\documentclass{llncs}
\usepackage{graphicx}
\usepackage{amssymb}
\usepackage{amsmath}
\usepackage{array}
\usepackage{cite}
\usepackage{tikz}
\usepackage{ifthen}
\usepackage{listings}
\usepackage{lstautogobble}
\usepackage[verbatim]{lstfiracode}

\newcommand{\ifarrow}[1][]{\ifthenelse{\equal{#1}{}}{\rightarrow}{-\hspace{-4.22pt}[{#1}]\hspace{-5.6pt}\rightarrow}}
\newcommand{\semarrow}[1][]{\ifthenelse{\equal{#1}{}}{\Rightarrow}{=\hspace{-3.0pt}[{#1}]\hspace{-3pt}\Rightarrow}}

\newcommand{\senc}{\textit{senc}}
\newcommand{\sdec}{\textit{sdec}}
\newcommand{\skzero}{\textit{SK}_{0}}
\newcommand{\sk}[1]{\textit{SK}_{#1}}
\newcommand{\kdf}{\textrm{KDF}}
\newcommand{\iv}{\textrm{IV}}

\title{Symbolic verification of Apple's Find My location-tracking protocol}
\author{Vaishnavi Sundararajan\inst{1}\orcidID{0000-0002-5945-5208} \and Rithwik\inst{2}}
\institute{Indian Institute of Technology, Delhi \email{vaishnavi@cse.iitd.ac.in} \and Independent researcher \email{rithuuik@gmail.com}}

\begin{document}
\maketitle
\begin{abstract}
Tracking devices, while designed to help users find their belongings in case of loss/theft, bring in new questions about privacy and surveillance of not just their own users, but in the case of crowd-sourced location tracking, even that of others even orthogonally associated with these platforms. Apple's Find My is perhaps the most ubiquitous such system which can even locate devices which do not possess any cellular support or GPS, running on millions of devices worldwide. Apple claims that this system is private and secure, but the code is proprietary, and such claims have to be taken on faith. It is well known that even with perfect cryptographic guarantees, logical flaws might creep into protocols, and allow undesirable attacks. In this paper, we present a symbolic model of the Find My protocol, as well as a precise formal specification of desirable properties, and provide automated, machine-checkable proofs of these properties in the Tamarin prover. 
\end{abstract}

\section{Introduction}
\subsection{Find My: History and evolution}
Apple's first location-tracking service bundled with iPhones (called ``Find My iPhone'') was originally released in 2010, and made free for all users with the release of iCloud in 2011. For Macs, ``Find My Mac'' was added to OS X 10.7 Lion. In the following few years, Apple released a service called ``Find My Friends'', which allowed people to view the location of their friends via the iCloud.com website. Both Find My iPhone and Find My Friends were built into the iPhone and could not be uninstalled, as of iOS 9 (2015)~\cite{Hug15}. In 2019, with the release of iOS 13 and MacOS 10.15 Catalina, Apple uniformized the location-tracking services for various devices into one application called Find My. 

As of today, Find My comes natively installed on almost the entire suite of Apple devices (including desktops, laptops, watches, earbuds/headphones, AirTags), as well as some non-Apple Find My-enabled devices from Belkin etc.~\cite{AppleFindMy}. The proprietary location-tracking service uses a crowd-sourcing method to track even devices which have no internet access. A Find My-enabled device can communicate using Bluetooth with nearby Apple devices, which can, upon connecting to the internet, relay the locations of these nearby-but-offline devices to Apple's servers, allowing people to locate even their offline devices. This system is claimed by Apple to be immune to tracking, thanks to a nifty end-to-end encryption scheme based on a system of periodically-changed keys. 

In February 2022, Apple acknowledged explicit concerns about the malicious use of their AirTags, which are Find My-enabled tracking tokens with a very small footprint, and can potentially be used to track or stalk people unobtrusively~\cite{AppSt22}. Updates were rolled out by Apple to mitigate the issue without compromising on the intrinsic purpose of location tracking; some of these included alerts provided to users if an AirTag not associated with their iCloud account was seen in proximity to their phones over time and across different locations. However, in~\cite{MFBM21}, the authors showed that this did not account for cloned/tampered-with versions of the AirTags, and provided a formalization of the protocol and a privacy-preserving variation (proved using standard cryptographic indistinguishability games) using more complex cryptographic primitives like blind signatures, in~\cite{MFBM21}.

As of January 2025, Apple claims that more than 2.35 billion Apple devices are in active use worldwide~\cite{Apple25}, so any security vulnerability that affects even a small fraction of these would be cause for concern. However, there has been no publicly-available formal analysis of the security guarantees claimed by Apple regarding Find My, since the actual source code is not freely available. Some papers (like~\cite{HSKH21}) have managed to reverse engineer the possible code based on freely-available descriptions of the same, and present some informal but rigorous results. Our symbolic analysis builds on this work, and the formal models in~\cite{MFBM21, EBGHJ24}.

\subsection{Find My: high-level description~\cite{HSKH21}}
The setup starts by pairing two Apple devices associated with the same iCloud account. A device which loses proximity to its paired device broadcasts Bluetooth Low Energy (BLE) advertisements to neighbouring devices using a rolling system of key generation. Any nearby Apple device connected to the internet will upload encrypted location reports including timestamps of such devices to Apple's servers in the cloud. Owner devices can then retrieve these location reports from Apple's servers at will, allowing them to find the ``last-known'' location of their devices. The claim is that the encrypted location reports maintain confidentiality of precise device locations, and a rolling key generation system for encryption deters any cross-referencing of location reports to uniquely identify -- and violate the privacy of -- a particular device (or a particular owner), even by Apple themselves.

\subsection{Contribution}
In this paper, we present the first formal symbolic analysis of a reconstruction of Apple's Find My protocol, automated in the Tamarin prover. We pin down a precise threat model and obtain a reasonable abstraction of the Find My protocol. We go by the descriptions in~\cite{HSKH21, MFBM21}, which analyze the protocol in a cryptographic setting. Towards this abstraction, we make reasonable design choices regarding a logical abstraction by collating information from various sources, and also formalize some properties of interest. Finally, we provide machine-checked proofs of this verification in Tamarin, which can be extended by other researchers, should a better description or a new version of Find My be released. Such proofs complement any cryptographic analysis, and go a long way towards engendering public trust in the veracity of Apple's claims regarding the security of such a widespread protocol, which is crucial given its limited prior analysis. This analysis will also help other efforts towards the symbolic analysis of similar protocols, and the design choices which might work well for obtaining faithful abstractions.

\subsection{Organization of the paper}
In Section~\ref{sec:relwork}, we discuss related work with respect to location-finding protocols in general, and Find My in particular, as well as in the field of automated formal verification of security protocols. In Section~\ref{sec:findmydetails}, we flesh out the details of the working of Find My, primarily from the information found in~\cite{EBGHJ24, HSKH21, MFBM21}. Section~\ref{sec:tamarin} deals with the abstraction and modelling of the Find My protocol, and the design choices made in order to capture this protocol and the desired properties in the Tamarin prover. This section also contains a list of the properties proved about the operation of Find My. Finally, we conclude with a discussion about this process of symbolically modelling a real-world protocol and lay out possible lines of future work in Section~\ref{sec:disc}.

\section{Related work}\label{sec:relwork}
\subsection{Location-finding via Bluetooth}
Bluetooth finders, which are small devices that track items using Bluetooth Low Energy connections, and associated location-finding protocols abound. Perhaps the first Bluetooth finder that become quite popular was Tile, which managed to raise \$2.6M via crowdfunding for its tags and app~\cite{Lom13}. The market was quickly flooded with similar devices (Duet, TrackR, StickNFind, iTrack Easy, Zizai Tech Nut etc.~\cite{Perez14}). It should be mentioned that none of these devices claimed to preserve user privacy while tracking devices, even though this is a reasonable expectation of any system that tracks one's devices, especially through crowd-sourcing. Security researchers, through reverse-engineering and other means, found that most of them suffered from multiple kinds of vulnerabilities, even by way of merely maintaining confidentiality of data, and also came up with new systems that could ensure privacy~\cite{Beards16, WCU20}. 

\subsection{Cryptographic analysis and claimed attacks}
As mentioned above, much analysis has been done for location-finding protocols, but not all of it has been formal. For Apple's Find My protocol,~\cite{HSKH21} pieces together publicly-available information as well as reverse-engineering techniques to recover a specification, and then performs some rigorous but empirical analysis to show attacks on location anonymity. (They also claim that Apple has fixed these vulnerabilities after they were responsibly disclosed to them before publication.) In~\cite{MFBM21}, the authors identify stalking attacks which can be effected using AirTags as well as AirTag clones, in spite of Apple's item safety alerts, which warn people if there is a strange AirTag in their vicinity for a sustained duration of time. They also present an improved version of Find My which uses a cryptographic primitive called partial blind signatures to ensure that such attacks are mitigated. The security for their ``Blind My'' protocol is formally proved using established cryptographic techniques of indistinguishability.

\subsection{Symbolic verification of real-world protocols}
However, it is well known that security protocols can admit logical attacks in spite of utilizing near-perfect cryptography (man-in-the-middle attacks, for instance), and one way to be assured of the absence of such attacks is to perform symbolic analysis of these systems. Symbolic analysis is orthogonal to cryptographic analysis, in that it assumes perfect cryptography upfront, and with an established attacker model, checks whether any logical flaws are possible by constructing an abstract representation of all possible executions of the system. Importantly, this is done without ever running the system. Symbolic analysis has been applied to many real-world systems in the recent past, including but not limited to authentication protocols like Kerberos~\cite{BCJ06}, network protocols like TLS~\cite{BFK13}, key exchange protocols like EDHOC~\cite{NSB21, NSB23}, and messaging protocols like Signal~\cite{KBB17}. Many automated tools have been developed and used which aid such verification; these include ProVerif~\cite{Bla16}, SAPIC+~\cite{CJKK22}, and Tamarin~\cite{MSCB13}. Symbolic analysis requires the casting of the system under investigation into an abstract, mathematical model, after which one also phrases the properties expected of this system's behaviour into the same kind of language. Finally, one checks whether this abstract model satisfies the formulated properties or not; if it does, one obtains a proof of correct (abstract) operation, and if it does not, one gets a counterexample by way of an offending execution. 

\section{Details of Find My, as in~\cite{HSKH21,MFBM21}}\label{sec:findmydetails}
There are four different algorithms involved in the operation of Find My. We will now describe each of these in detail. (Most of this is based on $\mathsection 6$ of~\cite{HSKH21} and $\mathsection 3$ of~\cite{MFBM21}.)


\subsection{Pairing algorithm}
In order to add a device $D$ to their iCloud account, an owner must communicate $D$'s serial number to Apple's servers. The server checks the validity of the serial number, and if it is invalid, the protocol aborts. Otherwise, a key establishment happens, whereby the owner device $O$ generates a private-public key pair of the form $(d_{0}, p_{0})$ using the NIST P-224 elliptic curve, and a symmetric key $\skzero$ of 32 bits. This information is securely communicated to $D$ since $O$ acts as a proxy to the internet for $D$, which requires that $D$ be close to the owner device at the time of pairing. Together, these form the \emph{master beacon key}. After pairing, the master beacon key is synchronized to iCloud in an encrypted file, the decryption key for which is stored in the iCloud keychain belonging to the owner. 

\subsection{Generating a beacon}
If $D$ senses that $O$ has moved away, $D$ is now ``lost'', and requires finding. The master beacon key is the only information that $D$ shares with $O$, and so the beacon message must be a (deterministic) function of this key. However, if this key is the only thing that uniquely identifies $O$ (and $D$, by extension, at $O$'s end), and every message utilizes this same key, privacy might be violated. In order to avoid such attacks on privacy, $D$ constructs beacons based on the concept of rolling keys, constructed using a key derivation function (more precisely, the ANSI X.963 KDF, with SHA-256, and a generator $G$ of the NIST P-224 curve).

The KDF operates in two modes: update and diversify. The update mode takes the symmetric key part of the current beacon, and generates a new symmetric key. This new symmetric key is then fed into the diversify mode of the KDF, which generates ``anti-tracking'' keys $u$ and $v$ of 36 bits each, which are used to generate the new private-public key pair, to finish the creation of the new beacon. The functions and equations to generate the new beacon (at the $i^{\rm th}$ ``epoch'') with symmetric key $\sk{i}$ and private-public key pair $(d_{i}, p_{i})$ are shown in Figure~\ref{fig:kdfeq}. (Here $*$ and $+$ are operations over the elliptic curve, $G$ is a generator, and $d_{0}$ is the private key part of the master beacon key.)
\begin{figure}
\vspace{-9mm}
\begin{minipage}{0.6\linewidth}
\begin{eqnarray}
\sk{i} &= \kdf(\sk{i-1}, \textrm{``update''}, 32) \label{eq:kdfeq1} \\
(u_{i}, v_{i}) &= \kdf(\sk{i}, \textrm{``diversify''}, 72) \label{eq:kdfeq2}
\end{eqnarray}
\end{minipage}
\begin{minipage}{0.35\linewidth}
\begin{eqnarray}
d_{i} = (d_{0}*u_{i}) + v_{i} \label{eq:kdfeq3} \\
p_{i} = d_{i} * G \label{eq:kdfeq4}
\end{eqnarray}
\end{minipage}
\vspace{-1mm}
\caption{Equations for the key-derivation function}
\label{fig:kdfeq}
\vspace{-5mm}
\end{figure}

During an epoch (of duration 15 minutes), devices emit one beacon every two seconds when they are in ``lost'' mode. Each beacon consists of the $p_{i}$ (as well as some bookkeeping metadata like BLE address of the device, proprietary headers etc.), i.e. the public part of the key generated for that epoch. The KDF as described above is a deterministic function, so the owner device can also do this computation at its own end to generate the new keys every epoch (and use the correct keys to decrypt later, upon receiving a encrypted location report ). If the elliptic curve is secure, there is no evident (to someone who does not know $G$ or the master beacon) dependence between successive beacons. 

\subsection{Creating and uploading location reports}
Suppose $D$ has been ``lost'', and some finder device $F$ is in proximity to $D$, and receives a BLE advertisement from $D$. The headers allow $F$ to parse this message as a Find My distress call. $D$ needs $F$ to upload a (confidential) location report witnessing a sighting of $D$ to Apple's servers such that nobody but the owner of $D$ can figure out the exact location of $D$ from this report. We revert to traditional methods of hiding information, namely encryption. However, the payload of $D$'s ad is a public key (which the owner of $D$ can also generate at their end, and is therefore assumed to have access to). One cannot securely encrypt using the public part of a private-public keypair. How can $F$ perform any encryption that can be reversed by $D$'s owner when they wish to locate $D$?

Recall that we use elliptic curve cryptography here, and therefore, one tool at our disposal is ephemeral Elliptic Curve Diffie-Hellman (ECDH) key exchange, which allows $F$ to derive a ``shared key'', which can be then used to encrypt the report. Note that the underlying choice of curve (NIST P-224) is known and the key derivation function $\kdf$ is accessible to all devices on the Find My network, including finder devices. Upon receiving a beacon containing a public key $p_{i}$, $F$ generates a new ephemeral key pair $(d_{f}, p_{f})$ on the curve (the keypair has to be freshly generated each time to avoid violating the privacy of $F$ itself). It then performs ECDH using $d_{f}$ and $p_{i}$ to generate a shared secret $ss$. One can derive a symmetric key by using the KDF on $ss$ and $p_{i}$, which yields a 32-byte message. The first 16 bytes are used as the encryption key $e'$, and the remaining 16 bytes are used as an initialization vector $\iv$. $F$ then encrypts its location using $e'$ and $\iv$ using a type of authenticated encryption with associated data (AES-GCM). Finally, the message that $F$ uploads to Apple's servers contains a timestamp, an ephemeral public key $p_{f}$, and the AES-GCM authentication tag (in the clear), as well as the encrypted location of $D$, and is stored using an ID which is the SHA-256 hash of the $p_{i}$ that was originally received in the BLE ad from $D$. 

\subsection{Accessing location reports}
Now suppose the owner $O$ of device $D$ realizes that $D$ is not close to them. Naturally, they wish to query for its last-known location, so they open the Find My application on their device. $O$ securely authenticates themselves to Apple's servers through their Apple ID. This allows them to download a list of ``recent'' location reports, indexed by metadata that includes the hashed values of the $p_{i}$ as well as the time at which the report was uploaded (not necessarily the same timestamp as inside the location report, since these reports are often uploaded in batches). 

\begin{figure}
\vspace{-2mm}
\begin{center}
\includegraphics[scale=0.21]{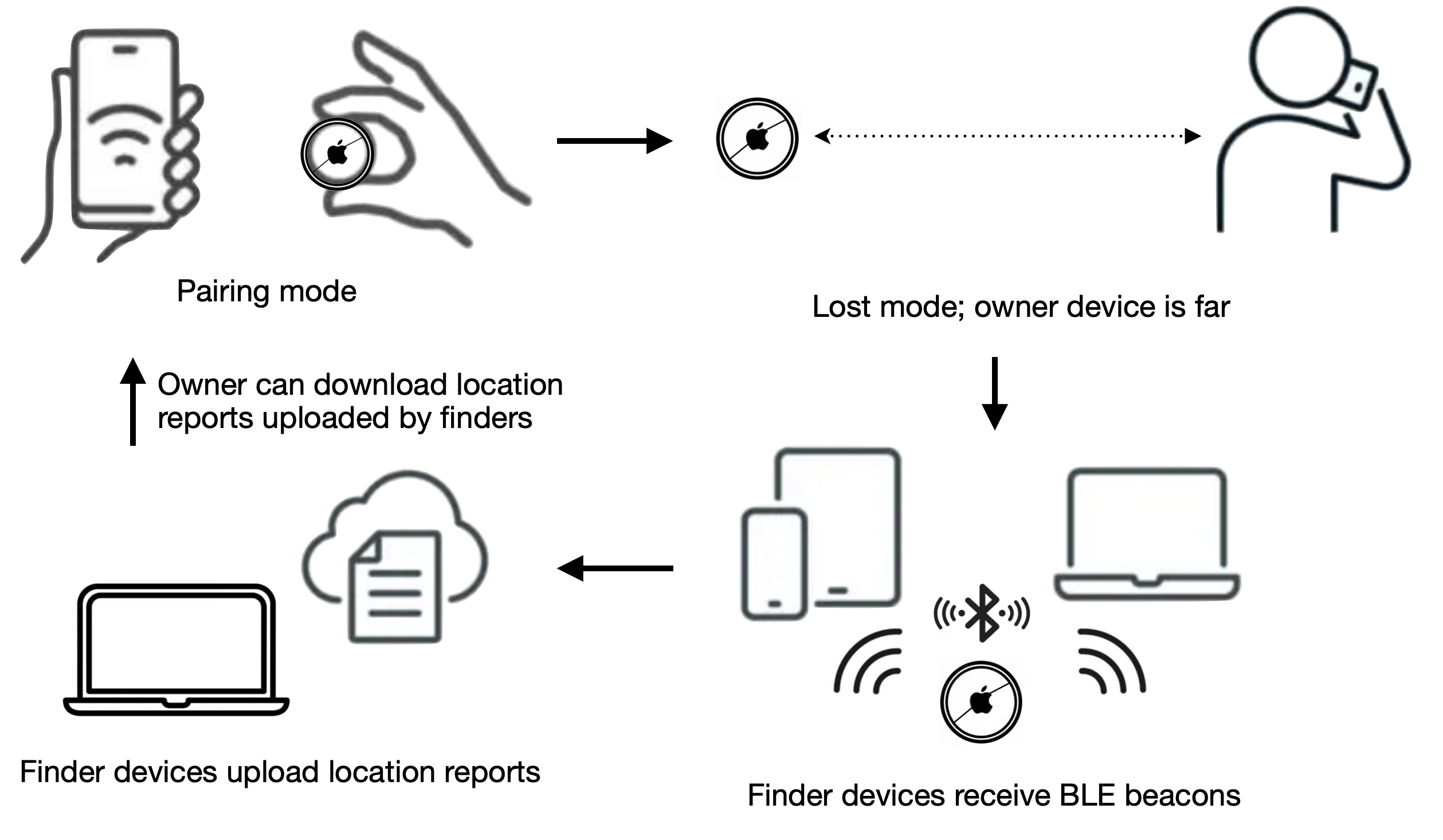}
\end{center}
\vspace{-2mm}
\caption{Find My Algorithms}
\label{fig:findmyworkflow}
\end{figure}

Given a location report $R$, using the hashed value of $p_{i}$ that comes as part of metadata, the owner $O$ runs through their list of $p_{i}$s for $D$, to compare which $p_{i}$ might have been used to construct this particular location report. Once the hash matches, the owner has access to the corresponding $d_{i}$ as well (since they know all pairs $(d_{i}, p_{i})$). Since an ECDH was performed using $d_{f}$ and $p_{i}$, and $p_{f}$ is sent in the clear as part of $R$, $O$ can generate the secret $ss$ (which is necessary to decrypt the actual location in $R$) by applying the same ECDH algorithm to $p_{f}$ and $d_{i}$. Once $O$ has $ss$, they can use $ss$ and $p_{i}$ to decrypt the encrypted location, and the Find My application combines the recent reports in the list to display the last-known location of $D$ on a map interface. The overall flow is shown in Figure~\ref{fig:findmyworkflow}.

\section{Formalizing Find My in Tamarin}\label{sec:tamarin}

\subsection{Introduction to Tamarin}
In this work, we use Tamarin~\cite{MSCB13} for the automated interactive symbolic verification of the Find My protocol. Tamarin operates over the standard Dolev-Yao model~\cite{DY83}, where all communicated messages are modelled as terms in an algebra. In Tamarin, protocols are written using multiset rewrite rules, which are stated over multisets of ``facts'', an aspect of the global system state. In this way, the multiset rewrite rules encode a transition system whose states stand for the global system state. A rewrite rule is of the form $l \ifarrow[e] r$, where $l$ and $r$ are multisets of facts, and the optional $e$ is a multiset of events, using which this rule can be annotated. Event annotations, called actions in Tamarin, allow us to state properties which talk about when a particular rewrite rule has been fired. Facts and actions are modelled as predicates of arbitrary arity (which also form their types), and can (and often will) mention and be parametrized using terms. The property language is a fragment of first order logic equipped with timestamps, and these statements are verified over system executions. Timestamps respect the usual quasi order, and multiple events can occur at the same time point, but with a type restriction: two events of the same type cannot happen at the same time instant.

Verification in Tamarin occurs in the presence of an equational theory, often represented by $E$. For example, in order to state that the $\sdec$ decryption operation using a key reverses the effect of the $\senc$ symmetric encryption operation with the same key, the following equation comes into play: $\sdec(\senc(u, v), v) =_{E} u$. This is different from an inference theory based approach, where the only function symbols in the algebra are the ``constructors'', which can build bigger terms from small ones, and inference rules govern how terms can be broken down (instead of having an $\sdec$ function, one would have an inference rule which captures the effect of $\sdec$). Tamarin has a default term algebra and an associated equational theory, but users are free to add new function symbols and equations for these additions. The equational theory is, therefore, fixed upfront, before the protocol is specified. For the purposes of this work, we will use Tamarin's built-in equational theories for symmetric and asymmetric encryption and hashing. In addition, we will also need some new functions and associated equations, which we will describe in detail in Section~\ref{sec:findmytamarin}.

\subsection{Threat model for Find My}\label{sec:findmythreatmodel}
We consider a standard threat model in symbolic verification, namely, that the network is assumed to be malicious, but cryptography is assumed to be perfect. This model was first proposed in~\cite{DY83}, and has held up remarkably well for modelling protocols even in today's day and age. Communicated messages are modelled as terms in an algebra, and it is assumed that the adversary cannot break cryptography, i.e. encrypted messages can only be decrypted with the correct decryption key, hashes are non-invertible and do not collide, signatures cannot be forged etc. The adversary is also assumed to control the network, and can observe all messages communicated as part of the protocol. Not only can they see all messages, they are allowed to interact maliciously with the flow of messages unless over a secure channel -- so the adversary can block, modify, inject, and redirect messages at will, they can masquerade as legitimate parties, and they can start new sessions with legitimate parties.

We also allow the adversary to access long-term and ephemeral keys via specific events, as in~\cite{NSB21, NSB23}. This allows us more precise control over what is possible if an adversary gains access to even these keys, which they might not in the course of ``normal'' operation. In addition, it also allows us to model collusion between (or a hostile takeover of) legitimate parties and the malicious attacker.

\subsection{Specifying the protocol in Tamarin}\label{sec:findmytamarin}
We present the code for specifying the Find My protocol (as described in Section~\ref{sec:findmydetails}) in this section. The full code can be found at~\cite{tamarinrepo}. Any departures in the snippets here from the code at that link are purely of a syntactic nature, or driven by a view to make the explanation simpler; there is no functional difference whatsoever in the verification thereof.

\vspace{-3mm}
\subsubsection{Initial setup and bookkeeping: }
We consider four parties: an owner $O$, a Find My-enabled device $L$ (often called an LTA, standing for ``location tracking accessory'', as in~\cite{EBGHJ24}, but the $L$ might as well stand for ``lost device''), a finder $F$ in proximity to $L$, and Apple's server $S$. Note an interesting property of an Apple device -- it can be an owner, an LTA, or a finder, depending on context! In Tamarin, facts come in two flavours, temporary and persistent. Temporary facts get ``consumed'' when a rewrite rule involving them gets fired; if a fact $f$ belongs to the multiset $l$ and a rewrite rule of the form $l \ifarrow[e] r$ is fired, $f$ is no longer available for some other rule which uses $f$ as a precondition to be fired (unless $f$ is generated again as part of $r'$ under some other rewrite rule $l' \ifarrow r'$). All facts are temporary by default. Permanent facts are those which do not get consumed, and are always available to be used as preconditions. A permanent fact in Tamarin is indicated by a ! preceding it. So while we start our protocol by assigning the owner role to the agent with name $O$, the LTA role to the agent with name $L$, and the finder role to the agent with name $F$, these facts are not permanent ones, since these same agents/devices could switch roles later. We indicate the fact that the agent with name $O$ is playing the owner role by way of a rule as follows.

\vspace{-1mm}
{
  \bgroup
  \begin{verbatim}
  rule Get_Owner: [] --> [ Owner($O) ]
  \end{verbatim}
  \egroup
}

\vspace{-4mm}
One reads this rewrite rule as having the name \verb|Get_Owner|, and requiring no prerequisites to be fired (hence the empty braces in the $l$ part of the rewrite rule), and upon successful execution, contributing the fact \verb|Owner($O)|. Any subsequent rules may use \verb|Owner($O)| as a precondition, but only one such rule will be fired, since this fact would be consumed by that rule. We add similar rules  for $L$, and $F$, while for $S$, since the Apple servers do not function as owners/lost devices/finders,  the rule contributes a persistent fact, so upon successful execution the fact that gets added is \verb|!Server($S)|.
\vspace{-3mm}
\subsubsection{Pairing phase: }
As mentioned earlier, during the pairing phase, the communication between $O$ and $L$ (not yet in ``lost mode'') is assumed to be implicitly secure, so we do not model that message as part of the protocol. Instead, we model the fact that if $d_{0}$ and $\skzero$ are ``freshly'' generated (we will see later how to enforce the ECDH properties for such values, to ensure that they are indeed generated from the elliptic curve properly), then both the owner and the LTA have access to them upfront, through the rule \texttt{GenKeys}, which looks as follows. The event annotation states that $d_{0}$ and $\skzero$ are established between $O$ and $L$.

{
  \bgroup
\begin{verbatim}
rule GenKeys: [ Fr(~d0), Fr(~SK0), Owner($O), LTA($L) ]
							--[ KeyEst($O, $L, ~d0, ~SK0) ]-> 
					[ Okd($O, $L, ~d0, ~SK0), Lkd($L, $O, ~d0, ~SK0) ]
\end{verbatim}
\egroup
}

\texttt{Fr} is a keyword in Tamarin, which enforces that the value assigned to the name therein (whose type is enforced with the preceding $\sim$) is freshly-generated, and does not match any other value that has been used. The rule says that if $d_{0}$ and $\skzero$ have been freshly generated, and if the agent with name $O$ is playing the role of an owner (\$ in Tamarin is a type signifier for agent names) and the agent with name $L$ that of an LTA, then $O$ and $L$ are assumed to have been paired with each other, indicated by the facts \texttt{Okd} and \texttt{Lkd}, parametrized by the agent names as well as the private key and the symmetric key. (The public part of the private-public keypair can be automatically derived using the elliptic curve, so we omit giving it an explicit name here.) This rule is all that is required to model the pairing phase of the Find My operation. In the other rules, we will omit type signifiers when not required. Here, we assume that the server is not storing the master beacon key in the clear, i.e. the iCloud keychain is not compromised.

\subsubsection{Granular key reveal: }
Once we have this rule to set up the keys, we also set up two rules which allow us finer access to when certain properties can be violated, namely the leak of the private portions of the master beacon key to the malicious adversary. This will allow us to model an adversary who can control/collude with $O$ or $L$, and gain access to the secret initial setup information held only by $O$, and allow us to verify whether the properties of interest hold even under such a strengthened threat model. This set of rules needs to be annotated with actions, so that we can refer to the firing of this rule in our lemmas when we model properties. The rule for the reveal of $d_{0}$ looks as follows (the rule for the reveal of $\skzero$ is very similar).

\vspace{-2mm}
\begin{verbatim}
rule Reveal_d0:
[ Lkd(L, O, d0, SK0) ] --[ LtkReveal_d0(O, L, d0) ]-> [ Out(d0) ]
\end{verbatim}

The rule is set up to fire only after the pairing phase is done (because what meaning would leaking the master beacon key have if it has not even been successfully established?), and hence requires the \texttt{Lkd} fact as a precondition. The rule uses the \texttt{Lkd} fact to say that if for some value of $L, O, d_{0}$, and $\skzero$, $L$ has been established as an LTA and paired with $O$ using $d_{0}$ and $\skzero$, then this rule \verb|Reveal_d0| is fired, and the $d_{0}$ key is sent out onto the channel (indicated by the keyword \texttt{Out}). Since under the Dolev-Yao model, one assumes that the adversary controls the network, this essentially leaks $d_{0}$ to the adversary. One can get evidence of this rewrite rule having been fired via the event \verb|LtkReveal_d0|, which also tells us which owner and which LTA were affected, and which private key was leaked. 

\vspace{-3mm}
\subsubsection{Lost phase (LTA operation): }
Now that we have set up all the initial bookkeeping, we are ready to proceed with specifying the more involved part of the protocol, i.e. the operations of the various parties when the LTA goes into lost mode. Lost mode communications depend on the exact choice of epoch, since the rotating keys are generated in a dovetailed fashion, taking the current epoch number ($i \ge 1$ in the equations in Figure~\ref{fig:kdfeq}) as a(n implicit) parameter. We model this rotation as an inductive definition: the base case is the one for the first epoch, and in the inductive step, we handle epoch number $i+1$ for arbitrary $i$. We could not explicitly set up a mechanism to churn out $d_{i}, p_{i}$ and $\sk{i}$ (we needed a global list to be dynamically populated but not accessible by all parties) and $d_{0}, p_{0}$ and $\skzero$ were freshly generated (without any ``seed'', as such), so the facts spawned are different in the initial setup versus any later iteration, so the function would anyway have been asymmetrical along that axis. Thus, we specify two separate rules for the LTA when in lost mode -- one for the base case, and one for the inductive case. We first describe the rule \verb|L_1| in Figure~\ref{fig:rulel1}, which describes the base case, where $d_{0}$ and $\skzero$ form the current beacon, and a new one is generated. We use the \texttt{let ... in} construct allowed by Tamarin to specify some local names which we will use in the specification of the rewrite rule that follows. 

\begin{figure}
\begin{verbatim}
rule L_1:
	let
	  SK_1 = SK_fn(~SK0)
	  d_1 = di_fn(~d0, SK_1)
	in
	[ Lkd($L, $O, ~d0, ~SK0) ]
	--[ LPFS1($L, $O, ~d0, ~SK0, d_1, SK_1),  Ok_s($L, $O, ~d0, ~SK0) ]->
	      [ Out(pk(d_1)), Fin_1(pk(d_1)), L_2($O, $L, ~d0, ~SK0, SK_1) ]
\end{verbatim}
\vspace{-5mm}
\caption{Rule for the behaviour of the lost device in epoch number 1}
\label{fig:rulel1}
\vspace{-5mm}
\end{figure}

We specify the KDF in two parts. The unary function \verb|SK_fn| captures the effect of equation~(\ref{eq:kdfeq1}) from Figure~\ref{fig:kdfeq}, and generates $\sk{1}$, given the current value of the symmetric key, which is $\sk{0}$. We abstract away the constant inputs to the function definition in equation~(\ref{eq:kdfeq1}). So essentially one can think of it as the following equation, converting this previously-ternary application of the $\kdf$ function into a unary definition of \verb|SK_fn|, which absorbs the effects of the constants ``update'' and 32: $\texttt{SK\_fn}(x) = \kdf(x, \textrm{``update''}, 32)$.

Similarly, the effect of the equations~(\ref{eq:kdfeq2}) and (\ref{eq:kdfeq3}) can also be captured in one single equation. Apart from $d_{0}$ and $\sk{i}$, the other arguments to the computation of $d_{i}$ are all constants. So we can abstract them out similar to the specification of \verb|SK_fn|, and write a binary function \verb|di_fn|. \vspace{-1.5mm}
\[ \texttt{di\_fn}(x, y) = (x*\textrm{fst}(\kdf(y, \textrm{``diversify''}, 72))) + \textrm{snd}(\kdf(y, \textrm{``diversify''}, 72)) \] 

\vspace{-1mm}
The prerequisites, in the first epoch, are merely that $L$ is paired as an LTA with some owner $O$, using $d_{0}$ and $\skzero$ (used in the computations for \verb|SK_1| and \verb|d_1|) as the master beacon key. This is witnessed by the fact \texttt{Lkd} with the appropriate parameters. If the rule fires, the effect is that the public part of the new beacon (expressed by applying the \texttt{pk} function to $d_{1}$, since these are the public and private components of an asymmetric key pair) is sent out onto the network. In addition, we add two new facts to the knowledge base: one that the public part of the new beacon is $\texttt{pk}(d_{1})$, and another that the new symmetric key generated as part of the rotation is $\sk{1}$. The event annotation for this rule deserves some explanation. \texttt{LPFS1} establishes that the LTA $L$ paired with the owner $O$ with master beacon parameters $d_{0}$ and $\skzero$ broadcast a BLE advertisement using the new beacon parameters $d_{1}$ and $\sk{1}$. \verb|Ok_s| is an annotation we add to ensure that there is at least one valid execution of the protocol, and our abstraction has not accidentally rendered the end result void.

The rule for the functioning of the lost device in any arbitrary epoch (other than epoch number 1) is very similar in structure, except that instead of using \texttt{Lkd} as a prerequisite, it uses \verb|L_2|, which should have been established by the operation of the lost device in epoch number 1 (i.e. by the successful firing of the rule \verb|L_1|). Similarly, the event annotation here has a different name, \texttt{LPFS2}. The sanity annotation is not necessary, since the first possible valid run would be captured in epoch number 1 itself (our helper lemmas tie any later execution to the execution of the rule \verb|L_1|).

\vspace{-3mm}
\subsubsection{Lost phase (Finder device operation): }
When a finder device receives a BLE ad from a lost device, it extracts the public key $p_{i}$ from the ad, and then runs the ECDH algorithm to generate an ephemeral keypair using which to encrypt the location. We first define a function \verb|SS_fn|, which generates the shared secret \texttt{SS} from a freshly-generated $d_{f}$ and $p_{i}$. Dual to what we did in the earlier section, where we captured the effect of a multi-ary function using a unary/binary function, here we split the KDF into two separate functions, both of which take as input \texttt{SS} and $p_{i}$. One of these (\texttt{KeyGen}) generates the first 16 bytes of the symmetric key, which is used as $e'$ (we name this \verb|e_p| in the code) to encrypt the location, and the other (\texttt{NonceGen}) generates the remaining 16 bytes, used as the initialization vector $\iv$. Recall that the encryption used for location reports is of the authenticated encryption with associated data kind, which requires us to define three new functions, namely \texttt{AEADenc}, \texttt{AEADauthdec}, and \verb|AEAD_dec|, along with the following equations to capture intended behaviour.

\vspace{-4mm}
\begin{eqnarray*}
\textrm{AEADauthdec}(k, \textrm{AEADenc}(k, pt, aad), aad) = pt \label{eq:aeadeqn1} \\ 
\textrm{AEAD\_dec}(k, \textrm{AEADenc}(k, pt, aad)) = pt \label{eq:aeadeqn2}
\end{eqnarray*}
\textrm{AEADenc} takes as input a key, a plaintext, and some associated data, which can be used to authenticate the encryption. The first equation specifies an authenticated decryption function which succeeds only if the keys match (the decryption key is the same as the encryption key) and the associated data used to encrypt matches the given input. The second equation specifies decryption which succeeds as long as the keys match, and does not check the associated data.

The crucial preconditions for the rewrite rule for the finder device are that a BLE ad should have been broadcast by a lost device (the \verb|F_1| fact should hold about some $p_{i}$, as a result of executing rule \verb|L_1| or \verb|L_2|), and the finder device should have received said ad (indicated by the \verb|In| keyword in Tamarin). The remaining are about freshly generating the timestamp, location, $d_{f}$ etc. If the preconditions are met, the rule outputs a tuple containing three terms: an encrypted object (using \verb|AEADenc|) containing the location and timestamp, using $\iv$ as the associated data, the public key corresponding to the ephemeral keypair generated by the finder, and the hash of the public key inside the received beacon (tuples are indicated in Tamarin using $\langle ... \rangle$). The rule also adds a fact regarding the uploading of this report (specified using the ephemeral public key and the hash of $p_{i}$) to the server by the finder.

{
  \bgroup
  \small
\begin{verbatim}
rule F_1:
	let
	  SS = SS_fn(~d_f, p_i)
	  e_p = KeyGen(SS, p_i)
	  IV = NonceGen(SS, p_i)
	  L = senc(~loc, e_p)
	in
	[ Fin_1(p_i), In(p_i), Finder($F), !Server($S), Fr(~tF), Fr(~loc), Fr(~d_f) ]
	   --[Floc(~loc, ~d_f, p_i), Eq(SS, SS_fn(~d_f, p_i))]->
 [ Out(<AEADenc(e_p, (<L, ~tF>), IV), pk(~d_f), h(p_i)>), 
 	F_upload($F, $S, pk(~d_f), h(p_i)) ]
\end{verbatim}
\egroup
}

\vspace{-2mm}
Interestingly, not only do we have an event annotation \verb|Floc|, which witnesses that this location was uploaded to the server by the finder, we also have an event \verb|Eq|. This is a keyword in Tamarin, which asserts equality between two terms, and here we state it between the shared secret \texttt{SS} and the function application \verb|SS_fn(~d_f, p_i)|. This seems unnecessary when we note that we are setting \texttt{SS} to be exactly this term that we are stating it to be equal to, but there is a deeper reason for doing such an annotation. In particular, it is to establish the ECDH between $d_{f}$ and $p_{i}$ (and their corresponding counterparts in the keypairs). To ensure this, we use the event annotations provided by \verb|Eq|, and write the following global restriction, which enforces the ECDH relationship. This says that for any terms, if there are two facts stating equality between a given $x$ and the \verb|SS_fn| function applied to some $df$ and $\texttt{pk}(di)$, and between a given $y$ and \verb|SS_fn(di, pk(df))|, then it must be that $x$ and $y$ are themselves equal. Event occurrences always need to be timestamped in Tamarin's syntax, so \texttt{\#i} and \texttt{\#j} are timestamps (the \# is a type indicator) and \verb|e@ #i| denotes that event $e$ happened at timestamp $i$.

{
\vspace{-2mm}
  \bgroup
  \small
\begin{verbatim}
restriction sec: "All x y df di #i #j. 
(Eq(x, SS_fn(df, pk(di))) @ #i) & (Eq(y, SS_fn(di, pk(df))) @ #j)  
	==> x = y"
\end{verbatim}
\egroup
}

\vspace{-6mm}
\subsubsection{Lost phase (Server and owner operation): }
For the server operation, we require a precondition that some finder has uploaded a location report, upon which the server rule spawns a fact that it has received this particular location report. When an owner queries for a location report, we once again split this into two different rules -- one for the first epoch, and one for any subsequent epoch (just like we did for the lost device). The owner sends their identity and the hash of the public key for the appropriate epoch to the server, and the server sends back the location report corresponding to that hash. We omit these rules here, since they are fairly straightforward, but the interested reader can look up~\cite{tamarinrepo} for the full details.

\vspace{-2mm}
\subsection{Formalizing properties in Tamarin}\label{sec:propertiestamarin}
We prove a sanity lemma, which checks to see if there is at least one valid execution of the protocol, and ensures that we have not made any egregious errors during our formalization, as well as some other helper lemmas.  We formalize the following desirable properties of Find My:

\begin{itemize}
\item \textbf{Secrecy of the master beacon key:} We wish to verify that the private component of the master beacon key, namely $d_{0}$ and $\skzero$, should be known only to the owner and the LTA which establish them during the pairing phase. We establish this via two separate lemmas, one of which establishes the secrecy of $d_{0}$, and the other the secrecy of $\skzero$. As mentioned earlier, in rule \verb|L_1|, we use the event annotation \texttt{Ld} to denote the fact that the rule has fired in the first epoch. We use this rule as an indicator of the fact that a lost device has used this value of $d_{0}$ to send out a beacon, and then state the lemma as ``If a lost device uses $d_{0}$ to create a beacon in the first epoch, then the adversary must not know $d_{0}$ unless they have performed a key reveal on $d_{0}$''. Tamarin's keyword \texttt{K} allows one to talk about adversary knowledge, and the key reveal rules we added earlier allow us the fine-grained control to talk about the reveal of $d_{0}$ upfront. As earlier, event occurrences need to be timestamped, so the lemma looks as follows.

\vspace{-2mm}
{
  \bgroup
  \small
\begin{verbatim}
lemma d0_sec:
" All O L d0 #i. Ld(L, O, d0) @ #i ==> 
		(not (Ex #j. K(d0) @ #j)) | (Ex #k. LtkReveal_d0(O, L, d0) @ #k) "
\end{verbatim}
\egroup
}

\vspace{-2mm}
We state a similar lemma for the secrecy of $\skzero$, using \texttt{LSK} instead of \texttt{Ld}. \\

\item \textbf{Secrecy of intermediate beacon key:} We write similar lemmas to verify the secrecy of any $d_{i}$ and $\sk{i}$, generated in any intermediate epoch. Note that here the precondition is slightly stronger to ensure termination, in that in addition to asking for the fact \texttt{LPFS2}, we also have to link the fact that \texttt{LPFS2} could only have happened after \texttt{Ld} and \texttt{LSK} (which must have happened earlier). Also, if the adversary manages to learn $d_{i}$, they must have access to both $d_{0}$ and $\skzero$. So the lemma for the secrecy of $d_{i}$ looks as follows.
{
  \bgroup
  \small
\vspace{-2mm}
\begin{verbatim}
lemma di_sec:
"All O L d0 SK0 d_i  Ski #i #j #k. 
LPFS2(L, O, d0, SK0, d_i, SKi) @ #i & Ld(L, O, d0) @ #j 
& LSK(L, O, SK0) @ #k & j < i & k < i ==>
(not (Ex #j. K(d_i) @ #j)) 
| ((Ex #k. LtkReveal_d0(O, L, d0) @ #k) 
& (Ex #k. LtkReveal_SK0(O, L, SK0) @ #k)) "
\end{verbatim}
\egroup
}

\item \textbf{Perfect forward secrecy (PFS) of master beacon key:} For the $d$ values in the beacon, this lemma states that if $d_{1}$ has been broadcast, and some $d_{i+1}$ has also been broadcast (in some appropriate future epoch), even if the value of $d_{1}$ is leaked to the adversary, they cannot figure out the value of $d_{i+1}$, unless they know $d_{0}$ and $\skzero$ (and therefore possess the power to reconstruct every future beacon by themselves). A similar lemma is stated for the $\sk{}$ values. \\

\item \textbf{Perfect forward secrecy (PFS) of intermediate beacon key:} For the $d$ values, we say that if $d_{1}$ has been broadcast, and some $d_{i}$ as well as $d_{i+j}$ have been broadcast, then even if the value of $d_{i}$ is leaked to the adversary, they should not be able to reconstruct the value of $d_{i+j}$ unless they know the master beacon key material. We state a similar lemma for the symmetric key. \\

\item \textbf{Secrecy of location:} The location value, even if communicated by the finder to the Apple server, and retrieved by the owner, should not be leaked. 

\end{itemize}

We coded up 12 lemmas to formalize these properties, including some helper lemmas. Of these, 10 were successfully verified, but the inductive-step versions of the secrecy and perfect forward secrecy of $\sk{i}$ timed out. This is interesting, and appears to be because the computation of $\sk{i}$ depends on the $\sk{i-1}$ computed in the previous iteration, and some well-foundedness appears to be failing, causing the proofs to go into a complicated infinite loop. We tried adding helper lemmas about epochs and $\sk{}$ as well as playing around with different heuristics, but that did not help. The complete results are given in Table~\ref{tab:tamarinresults} in Appendix~\ref{app:tabresults}.

\section{Discussion and future work}\label{sec:disc}
In this paper, we performed the first formal symbolic analysis of Apple's location-tracking protocol called Find My. We used the reconstructions of the protocol from~\cite{HSKH21, MFBM21} to arrive at a reasonable abstraction for the protocol. This was a difficult task because the source code is not freely available, and the reverse-engineering efforts also suffer from some drawbacks. The cryptographic abstraction in~\cite{MFBM21} abstracts away much of the actual working of the protocol, which is necessary for us to form a faithful symbolic abstraction, and the work in~\cite{HSKH21} is written to appeal to the practitioner/hacker, and does not provide precise-enough descriptions of what the contents of various messages are, or how they are generated. These papers are wildly different in tone, and gleaning a full picture of the protocol from these papers took a lot of time and work, especially to ensure that we did not smuggle in incorrect claims. 

Even after we had a somewhat clear picture of the way the Find My protocol works in its various modes at various end points, it was hard to model this accurately in Tamarin, and we had to make many design choices about how to handle the encryption scheme, what the associated data was, when an end-point's identity is known and when it is not. This required us to add a function and equality rules for authenticated encryption with associated data, to choose $\iv$ to be the associated data, as well as include the restriction on equality to handle the on-the-fly ECDH and restrict the search space in a sound manner. The inductive definition for epochs was difficult to come up with while remaining faithful to the protocol operation, and also required us to be very careful about stating our lemmas in a fashion such that we did not miss any attack cases. Tamarin has a notion of state, but the epoch number is not globally known, so we could not maintain a global state, nor could we synchronize individual agent states (since both the owner and lost device know what epoch it is, but nobody else should), without resorting to this inductive modelling. In particular, this doubled our work, since we now needed to state one lemma for the initial case (epoch number 1) and one for the others -- and in some cases, the lemma for the arbitrary epoch with index $> 1$ was significantly harder to get to terminate. This is also an artefact of using Tamarin, in that some particular formulations of lemmas can be quite difficult to ensure termination for in Tamarin (as for the inductive cases for $\sk{}$ for us). We leave as future work the design of specialized oracles which might help termination in such ornery cases. 

One aspect of using Tamarin (or most other symbolic verification tools for security protocols) is that agents are often uniquely identified using their keys, since the adversary (under the Dolev-Yao model) is assumed to be able to impersonate anyone, and the standard way to talk about malicious collusion/control is to leak an agent's key to the attacker. In this particular modelling, we have no specific key for the Apple server, but we do wish to state that the server stores all the master beacon keys (via the keychain) and identification information to compare against when devices upload or retrieve location reports. This could be misused if the server itself is assumed to be malicious. Currently we can model part of a malicious operation by a server saying that if the master beacon key is known (which the server knows) then certain secrecy and perfect forward secrecy properties fail. However, we cannot capture the full gamut of a malicious server without having access to a specific key that also appears as part of the protocol, to pinpoint the exact footprint of such a malicious server. The same issue also crops up for the finder devices.

Another artefact of this choice of tool was that we did not replicate the indistinguishability properties given in~\cite{MFBM21}. Indistinguishability properties fall under the class of equivalence properties, which means that verifying such properties requires the simultaneous examination of multiple executions of the system. Tamarin has a ``diff'' mode, where one can perform such verification, but it is manual and hard to use, as compared to a tool like ProVerif, where such verification (while slightly more restricted) has a more plug-and-play flavour. Many properties of interest, unfortunately, fall into this class. Privacy/indistinguishability of devices involved in the Find My protocol is a major equivalence property, and we would like to verify this in the symbolic model as well, like~\cite{MFBM21} has done for some properties in the cryptographic model. Currently it is not possible to, since everyone is assumed to know every agent's name/identifier, and verifying a lemma like ``If $O$ retrieves location reports from an Apple server, the adversary should not know $O$'s name'' will trivially fail. Such properties we relegate to future work, perhaps modelling the same protocol in a different tool. This would also give us a clear sense of what design choices are forced by our choice of tool, and whether we can relax these somewhat. We also intend to investigate the security claims of Find My under a PQC model, along the lines of the work in~\cite{BCZ23}, in the future.

\bibliographystyle{plain}
\bibliography{ref.bib}

\begin{thebibliography}{10}

\bibitem{AppSt22}
{Apple Newsroom}.
\newblock {An Update on AirTag and Unwanted Tracking}.
\newblock \url{https://nr.apple.com/d2I9N827r9}.
\newblock Accessed: 2025-10-10.

\bibitem{Apple25}
{Apple Newsroom}.
\newblock {Apple Reports First Quarter Results}.
\newblock
  \url{https://www.apple.com/newsroom/2025/01/apple-reports-first-quarter-results/}.
\newblock Accessed: 2025-10-10.

\bibitem{BFK13}
Karthikeyan Bhargavan, C{\'e}dric Fournet, Markulf Kohlweiss, Alfredo Pironti,
  and Pierre-Yves Strub.
\newblock Implementing {TLS} with verified cryptographic security.
\newblock In {\em 2013 IEEE Symposium on Security and Privacy}, pages 445--459.
  IEEE, 2013.

\bibitem{BCZ23}
Nina Bindel, Cas Cremers, and Mang Zhao.
\newblock {FIDO2}, {CTAP 2.1}, and {WebAuthn 2}: Provable security and
  post-quantum instantiation.
\newblock In {\em 2023 IEEE Symposium on Security and Privacy (SP)}, pages
  1471--1490. IEEE, 2023.

\bibitem{Bla16}
Bruno Blanchet.
\newblock {Modeling and verifying security protocols with the applied pi
  calculus and ProVerif}.
\newblock {\em Foundations and Trends in Privacy and Security}, 1(1):1--135,
  2016.

\bibitem{BCJ06}
Frederick Butler, Iliano Cervesato, Aaron~D Jaggard, Andre Scedrov, and
  Christopher Walstad.
\newblock Formal analysis of {K}erberos 5.
\newblock {\em Theoretical Computer Science}, 367(1-2):57--87, 2006.

\bibitem{CJKK22}
Vincent Cheval, Charlie Jacomme, Steve Kremer, and Robert K{\"u}nnemann.
\newblock $\{$SAPIC+$\}$: Protocol verifiers of the world, unite!
\newblock In {\em 31st USENIX Security Symposium (USENIX Security 22)}, pages
  3935--3952, 2022.

\bibitem{DY83}
Danny Dolev and Andrew Yao.
\newblock {On the security of public-key protocols}.
\newblock {\em IEEE Transactions on Information Theory}, 29(2):198--208, 1983.

\bibitem{EBGHJ24}
Harry Eldridge, Gabrielle Beck, Matthew Green, Nadia Heninger, and Abhishek
  Jain.
\newblock Abuse-resistant location tracking: Balancing privacy and safety in
  the offline finding ecosystem.
\newblock In {\em 33rd USENIX Security Symposium (USENIX Security 24)}, pages
  5431--5448, 2024.

\bibitem{HSKH21}
Alexander Heinrich, Milan Stute, Tim Kornhuber, and Matthias Hollick.
\newblock {Who Can Find My Devices? Security and Privacy of Apple’s
  Crowd-Sourced Bluetooth Location Tracking System}.
\newblock {\em Proceedings on Privacy Enhancing Technologies}, 3:227--245,
  2021.

\bibitem{AppleFindMy}
{iCloud User Guide}.
\newblock {What you can locate with Find My on each device}.
\newblock
  \url{https://support.apple.com/en-in/guide/icloud/mm82e8ac5129/icloud}.
\newblock Accessed: 2025-10-10.

\bibitem{KBB17}
Nadim Kobeissi, Karthikeyan Bhargavan, and Bruno Blanchet.
\newblock Automated verification for secure messaging protocols and their
  implementations: A symbolic and computational approach.
\newblock In {\em 2017 IEEE European symposium on security and privacy
  (EuroS\&P)}, pages 435--450. IEEE, 2017.

\bibitem{MFBM21}
Travis Mayberry, Ellis Fenske, Dane Brown, Jeremy Martin, Christine Fossaceca,
  Erik~C Rye, Sam Teplov, and Lucas Foppe.
\newblock {Who Tracks the Trackers? Circumventing Apple's Anti-Tracking Alerts
  in the Find My Network}.
\newblock In {\em Proceedings of the 20th Workshop on Workshop on Privacy in
  the Electronic Society}, pages 181--186, 2021.

\bibitem{MSCB13}
Simon Meier, Benedikt Schmidt, Cas Cremers, and David Basin.
\newblock {The TAMARIN prover for the symbolic analysis of security protocols}.
\newblock In {\em 25th International Conference on Computer Aided
  Verification}, volume 8044 of {\em Lecture Notes in Computer Science}, pages
  696--701, 2013.

\bibitem{Lom13}
TechCrunch {Natasha Lomas}.
\newblock {Tile Grabs \$2.6M Via Selfstarter For Its Lost Property-Finding
  Bluetooth Tags Plus App}.
\newblock
  \url{https://techcrunch.com/2013/07/24/tile-grabs-2-6m-via-selfstarter-for-its-lost-property-finding-bluetooth-tags-plus-app/}.
\newblock Accessed: 2025-10-09.

\bibitem{Hug15}
{Neil Hughes, Apple Insider}.
\newblock {Find My iPhone, Friends become built-in apps in Apple's first iOS 9
  beta, cannot be uninstalled}.
\newblock
  \url{https://appleinsider.com/articles/15/06/10/find-my-iphone-friends-become-built-in-apps-in-apples-first-ios-9-beta-cannot-be-uninstalled}.
\newblock Accessed: 2025-10-10.

\bibitem{NSB21}
Karl Norrman, Vaishnavi Sundararajan, and Alessandro Bruni.
\newblock Formal analysis of {EDHOC} key establishment for constrained {IoT}
  devices.
\newblock In {\em SECRYPT 2021, ISBN 978-989-758-524-1}, pages 210--221, 2021.

\bibitem{NSB23}
Karl Norrman, Vaishnavi Sundararajan, and Alessandro Bruni.
\newblock Extended formal analysis of the {EDHOC} protocol in {T}amarin.
\newblock {\em E-Business and Telecommunications, Communications in Computer
  and Information Science}, 1795:224--248, 2023.

\bibitem{Perez14}
{Sarah Perez, TechCrunch}.
\newblock {Duet Takes On Tile With A Small, Square Lost Item Finder That Also
  Lets You Replace The Battery}.
\newblock
  \url{https://techcrunch.com/2014/04/23/duet-takes-on-tile-with-a-small-square-lost-item-finder-that-also-lets-your-replace-the-battery/}.
\newblock Accessed: 2025-10-09.

\bibitem{Beards16}
{Tod Beardsley, rapid7.com}.
\newblock {Multiple Bluetooth Low Energy (BLE) Tracker Vulnerabilities}.
\newblock
  \url{https://www.rapid7.com/blog/post/2016/10/25/multiple-bluetooth-low-energy-ble-tracker-vulnerabilities/}.
\newblock Accessed: 2025-10-09.

\bibitem{tamarinrepo}
{Vaishnavi Sundararajan, Rithwik}.
\newblock {Tamarin code for Find My symbolic verification}.
\newblock \url{https://tinyurl.com/tacas-findmy}.

\bibitem{WCU20}
Mira Weller, Jiska Classen, Fabian Ullrich, Denis Wa{\ss}mann, and Erik Tews.
\newblock {Lost and Found: Stopping Bluetooth Finders from Leaking Private
  Information}.
\newblock In {\em Proceedings of the 13th ACM Conference on Security and
  Privacy in Wireless and Mobile Networks}, pages 184--194, 2020.

\end{thebibliography}

\vfill
\pagebreak 

\appendix 

\section{Verification results}\label{app:tabresults}

\begin{center}
\begin{table}
\caption{Results of verification in Tamarin for the various lemmas}
\begin{tabular}{ | m{8em} | m{25em} | m{5em} | }
\hline 
 Lemma name & Description & Status \\
 \hline
 \verb|sanity_check| & Is the protocol executable at all? & Verified\\  
 \verb|epochs_start1| & If an LPFS1 event happens, a KeyEst event with the same parameters must have happened beforehand &  Verified \\
 \verb|epochs_start2| & If an LPFS2 event happens, an LPFS1 event must have happened beforehand between the same parties & Verified \\
 \verb|epochs_end| & If a finder receives a distress beacon, LPFS1 or LPFS2 must have happened with the same LTA beforehand & Verified \\\
 \verb|d0_sec| & $d_{0}$ is known only to the owner and the LTA who paired using it & Verified \\
 \verb|SK0_sec| & $\skzero$ is known only to the owner and the LTA who paired using it & Verified \\
 \verb|di_sec| & Any arbitrary $d_{i}$ (involved in an LPFS1 or LPFS2 event) must be known only to its owner and LTA & Verified \\
 \verb|ski_sec| & Any arbitrary $\sk{i}$ (involved in an LPFS1 or LPFS2 event) must be known only to its owner and LTA & Timed out \\
 \verb|pfs_init_d| & Even if $d_{1}$ (from the first epoch) is leaked, the adversary should not get to know any future $d_{i}$s & Verified \\
 \verb|pfs_d| & Even if an arbitrary $d_{i}$ is leaked, the adversary should not get to know any future $d_{i+j}$s & Verified \\
 \verb|pfs_init_sk| & Even if $\sk{1}$ (from the first epoch) is leaked, the adversary should not get to know any future $\sk{i}$s & Verified \\
 \verb|pfs_sk| & Even if an arbitrary $\sk{i}$ is leaked, the adversary should not get to know any future $\sk{i+j}$s & Timed out \\
 \hline
\end{tabular}
 \label{tab:tamarinresults}
\end{table}
\end{center}

\end{document}